\documentclass[a4paper]{article}

\usepackage{spconf}
\usepackage{amsmath,graphicx,tikz,amssymb,booktabs}
\usepackage{hyperref}
\usepackage{fancyhdr}
\hypersetup{hidelinks,}
\usetikzlibrary{backgrounds}
\usetikzlibrary{positioning}
\usetikzlibrary{calc}
\usetikzlibrary{shapes.multipart}
\usetikzlibrary{arrows.meta}
\usetikzlibrary{decorations.pathreplacing}
\usetikzlibrary{fit}

\fancypagestyle{firstpage}{%
\fancyhf{}
\fancyfoot[C]{\copyright2018 IEEE. Published in the IEEE 2018 Workshop on Spoken Language Technology (SLT 2018),\\scheduled for 18-21 December 2018 in Athens, Greece.}

}

\title{An exploration of mimic architectures \\ for residual network based spectral mapping}
\name{Peter Plantinga, Deblin Bagchi, Eric Fosler-Lussier}
\address{The Ohio State University}

\begin{document}

\maketitle
\thispagestyle{firstpage}
\begin{abstract}

Spectral mapping uses a deep neural network (DNN) to map directly from noisy speech to clean speech. 
%This system can be used as a generic front-end denoiser to any speech recognition pipeline. 
Our previous study~\cite{bagchi2018spectral} found that the performance of spectral mapping improves greatly when using helpful cues from an acoustic model trained on clean speech. The mapper network learns to mimic the input favored by the spectral classifier and cleans the features accordingly. In this study, we explore two new innovations: we replace a DNN-based spectral mapper with a residual network that is more attuned to the goal of predicting clean speech.  We also examine how integrating long term context in the mimic criterion (via wide-residual biLSTM networks) affects the performance of spectral mapping compared to DNNs. Our goal is to derive a model that can be used as a preprocessor for any recognition system; the features derived from our model are passed through the standard Kaldi ASR pipeline and achieve a WER of 9.3\%, which is the lowest recorded word error rate for CHiME-2 dataset using only feature adaptation. %Our system is devoid of customized pipelines, fancy language models and additional robust features.

\end{abstract}

\section{Introduction}
\label{sec:intro}

Applying deep learning to the task of Automatic Speech Recognition (ASR) has shown great progress recently in clean environments. However, these ASR systems still suffer from performance degradation in the presence of acoustic interference, such as additive noise and room reverberation.

One strategy to address this problem is to use a deep learning front-end for denoising the features, which are then fed to the ASR system. Some of these models attempt to estimate an ideal ratio mask (IRM) which is multiplied with the spectral features to remove noise from the speech signal \cite{narayanan2015improving}. Others utilize spectral mapping in the signal domain \cite{han2014learning,han2015learning} or in the feature domain \cite{han2015deep,bagchi2015combining} to translate directly from noisy to clean speech without additional constraints.

When these pre-processing models were introduced, they could be easily decoupled from the rest of the ASR pipeline. This was useful, because these models provided a general-purpose speech denoising module that could be applied to any noisy data. With time, impressive gains in performance were noticed with the addition of noise-robust features and joint training of spectral mapper and acoustic model~\cite{wang2016joint}. However, the front-end and back-end models in these approaches each depend on the presence of the other, i.e. one would not be able to re-use the mapper for another task or dataset without re-training it. Moreover, adding robust features increases the difficulty of feature creation and increases the number of parameters in the speech recognition pipeline.
% * <plantinga.peter@protonmail.com> 2018-07-05T15:53:17.339Z:
% 
% > Moreover, joint training significantly increases the number of parameters of a model making the model harder to converge.
% 
% Does it really, though? My understanding is that joint training doesn't increase parameter count at all. Adding noise-robust features might though.
% 
% 
% ^.
%An ideal solution would not be dependent on any particular acoustic model or dataset, so that it can be used with other ASR systems or even for other speech-related tasks. After training, this model could then be used as a general-purpose denoiser for many ASR pipelines.

\begin{figure}[t]
%\hspace{-0.3cm}
\tikzstyle{label} = [fill=white,opacity=0.7,text opacity=0.9,font=\small]
\centering
\begin{tikzpicture}
	
	\node[rectangle,draw] (time) at (0,0) {\includegraphics[width=3cm,height=1.7cm]{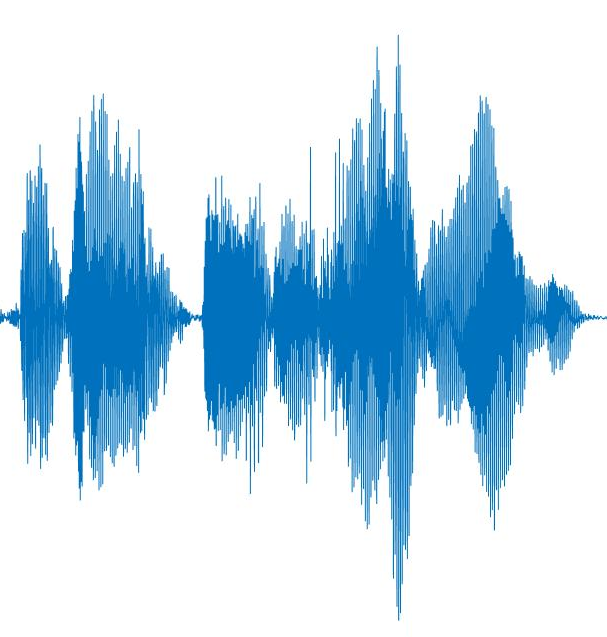}};
    \node[label,above=0.1em of time.south] (raw) {raw waveform};
    
    \node[rectangle,draw,below=3.5em of time] (noisy) {\includegraphics[width=3cm,height=1.7cm]{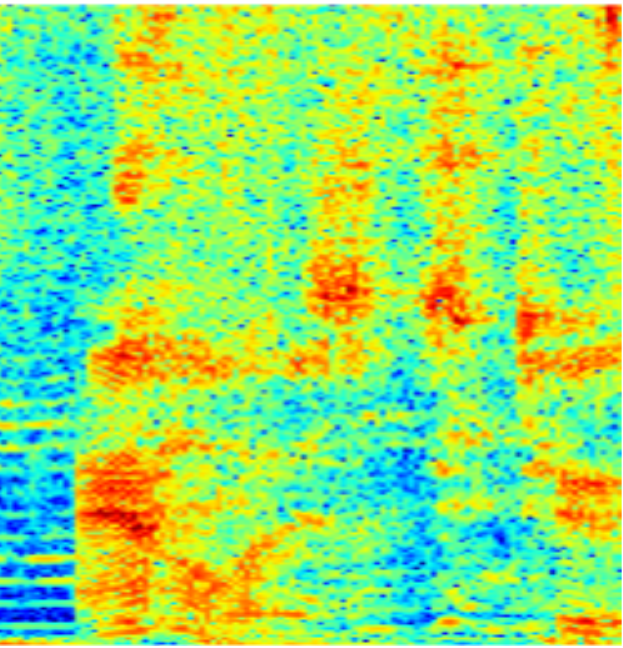}};
    \node[label,above=0.1em of noisy.south] (spec) {noisy spectrogram};
    
    \node[rectangle,draw,below=3.5em of noisy] (denoised) {\includegraphics[width=3cm,height=1.7cm]{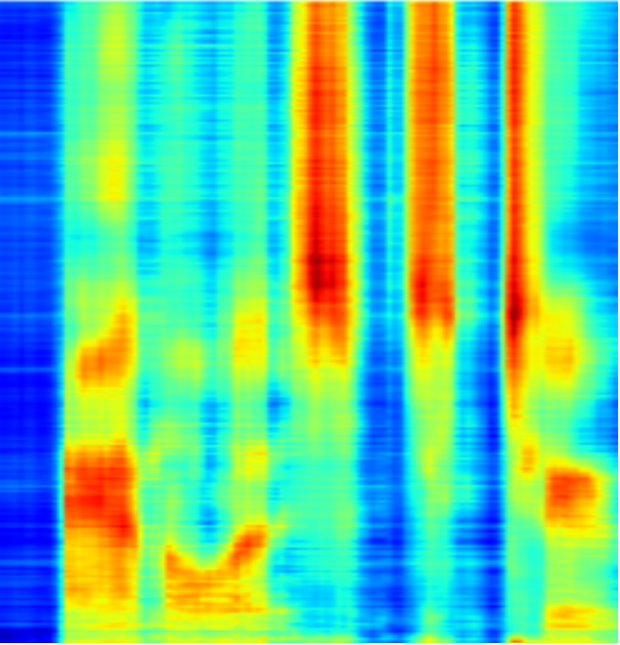}};
    \node[label,above=0.1em of denoised.south] (despec)  {cleaned spectrogram};
    
    \node[rectangle,draw,below=3.5em of denoised] (senones) {\includegraphics[width=3cm,height=1.7cm]{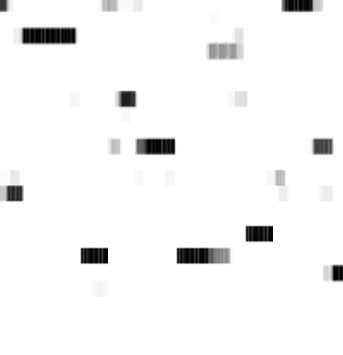}};
    \node[label,above=0.1em of senones.south] (post) {posteriorgram};
    %\node[above=2cm of time.north east] {\LARGE\textbf{Mapping Pipeline}};
    \draw[decoration={brace,raise=5pt,amplitude=0.3cm,aspect=0.1},decorate](time.north east) -- node[right=0.5cm,pos=0.1] {\Large{Step 1: Denoising}} (senones.south east);
    %\node[below=1em of time] (step1) {\LARGE{Step 1}};
    
    %\node[draw,color=red,inner sep=1cm,dashed,fit={(noisy.north) (senones.south west) (senones.south east)}] {};

    \node[rectangle,draw,right=1.1cm of noisy] (denoised2) {\includegraphics[width=3cm,height=1.7cm]{figures/denoised-nolabel}};
    \node[label,above=0.1em of denoised2.south] (despec2)  {cleaned spectrogram};
    \node[rectangle,draw,below=3.5em of denoised2] (senones2) {\includegraphics[width=3cm,height=1.7cm]{figures/posteriorgram}};
        \node[label,above=0.1em of senones2.south] (post2) {posteriorgram};
    \node[rectangle,draw,below=3.5em of senones2] (words) {\includegraphics[width=3cm,height=1.7cm]{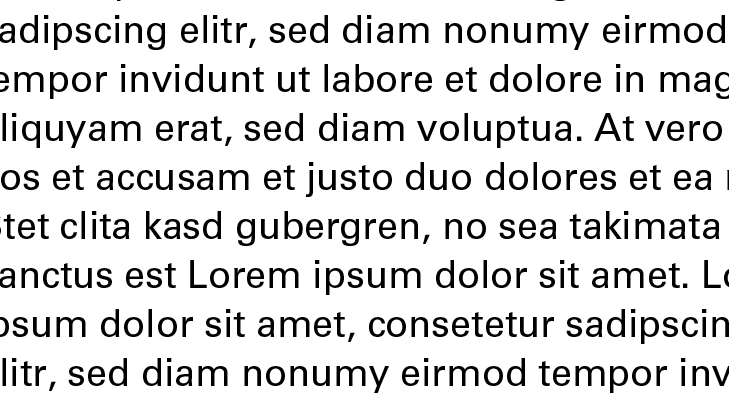}};
    \node[label,above=0.1em of words.south] (word) {recognized words};
    %\node[above=2em of denoised2] (label2) {\LARGE Speech Recognition};
    %\node[above=2em of label2] (label1) {\LARGE Feature Denoising};
    \draw[decoration={brace,raise=5pt,amplitude=0.3cm,aspect=0.5},decorate](denoised2.north west) -- node[above=0.5cm,pos=0.5] {\Large{Step 2: ASR}} (denoised2.north east);
    %\node[above=1em of words] (step2) {\LARGE{Step 2}};
    
    \draw[-stealth,thick] (time) -- node[left=0.4cm] {STFT} (noisy);
    
    \draw[-stealth,ultra thick] (noisy) -- node[left] {\textbf{\begin{tabular}{c}Spectral \\ mapping\end{tabular}}} (denoised);
    
    %\draw[arr] (noisy) -- node {\begin{tabular}{c}Spectral\\mapping\end{tabular}} (denoised2);
    
    \draw[-stealth,ultra thick,color=red,dashed] ([xshift=0.3cm] denoised.north) -- node[right=-0.1em] {\textbf{\begin{tabular}{c}Mimic\\loss\end{tabular}}} ([xshift=0.3cm] noisy.south);
    
    \draw[-stealth,thick] (denoised) -- node[left] {\begin{tabular}{c}Acoustic \\ modeling\end{tabular}} (senones);
    
    \draw[-stealth,thick,color=red,dashed] ([xshift=0.3cm] senones.north) -- node[right] {\begin{tabular}{c}Mimic \\ loss\end{tabular}} ([xshift=0.3cm] denoised.south);
    
    \draw[-stealth,ultra thick] (denoised2) -- node[right] {\textbf{\begin{tabular}{c}Acoustic\\modeling\end{tabular}}} (senones2);
    
    \draw[-stealth,thick] (senones2) -- node[right=0.2cm] {Decoding}  (words);
    
\end{tikzpicture}
\caption{System pipeline for spectral mapping with mimic loss. Bold text indicates training a model.}
\label{fig:pipeline}
\end{figure}

Our previous work~\cite{bagchi2018spectral} introduced a form of knowledge transfer we dubbed \textit{mimic loss}.  Unlike student-teacher learning \cite{ba2014deep} or knowledge distillation \cite{hinton2015distilling,lopez2015unifying,li2014learning} which transfer knowledge from a cumbersome model to a small model, the mimic approach transfers knowledge from a higher-level model (in this case, an acoustic model) to a lower-level model (a noisy to clean transformation). This can be seen in context in Figure \ref{fig:pipeline}. %seek to build a simpler model by imitating the soft outputs of a more complicated model, the mimic approach learns an input transformation from noisy to clean speech that makes a fixed phonetic (senone) classifier trained on clean speech behave similarly under noisy and clean speech.
% * <plantinga.peter@protonmail.com> 2018-07-06T22:13:48.670Z:
% 
% > a form of student-teacher learning
% My understanding was that mimic loss and student-teacher learning were both forms of knowledge distillation, but that mimic loss is not a form of student-teacher learning
% 
% ^ <deblinbagchi@gmail.com> 2018-07-16T14:43:50.585Z:
% 
% As far as I understand, Teacher-student learning and knowledge distillation are one and the same thing.  Ba and Caruana coined the term "teacher-student learning" in early 2000s (the paper we have cited in ICASSP) and Hinton  came up with "knowledge distillation" 2 years ago. The only difference is Hinton uses a KL-div loss on outputs whereas Caruana uses MSE loss on pre-softmax. Both Knowledge distillation and Teacher-student learning can be used on simple as well as "complex" models 
%
% ^ <fosler@cse.ohio-state.edu> 2018-07-16T14:56:27.657Z:
% 
% Maybe we should use "standard teacher-student learning" here?
%
% ^ <deblinbagchi@gmail.com> 2018-07-16T15:10:13.217Z:
% 
% this statement feels a bit controversial "Unlike student-teacher learning which seeks to build a simpler model imitating a more complicated model".....why don't we just start with "mimic approach...." ?
%
% ^ <plantinga.peter@protonmail.com> 2018-07-16T15:17:37.508Z:
% 
% Reading the Hinton paper, its a little more sophisticated than simply using cross-entropy on softmax outputs, because they play with the temperature quite a bit. But you're totally right that my categorization of student-teacher learning and mimic loss as both forms of knowledge distillation was incorrect.
%
% ^.
In this work, we improve our results using the mimic loss framework in two ways:

%\begin{enumerate}
First, we propose a residual network~\cite{he2016deep} for spectral mapping. A residual network model is a natural fit for the task of speech denoising, because like the model, the task involves computing a residual, i.e. the noise contained in the features. We find that a residual network architecture by itself works well for the task of speech enhancement, surpassing the performance of other front-end-only systems.
% * <fosler@cse.ohio-state.edu> 2018-07-06T01:06:11.120Z:
% 
% > residual network
% citation needed
% 
% ^ <plantinga.peter@protonmail.com> 2018-07-06T02:01:43.416Z.

Second, we use a more sophisticated architecture for senone classification, since this is the backbone of mimic loss. This provides a more informative error signal to the spectral mapper. To achieve this goal, we choose Wide Residual BiLSTM Networks (WRBN)~\cite{jahn2016wide} as the architecture for our senone classifier, which combines the effective feature extraction of residual networks~\cite{he2016deep} and the long-term context modeling of recurrent networks~\cite{graves2013speech,graves2013hybrid}.
% * <fosler@cse.ohio-state.edu> 2018-07-06T01:05:55.076Z:
% 
% > sequence discriminative modeling of recurrent networks
% citation needed here
% 
% ^ <plantinga.peter@protonmail.com> 2018-07-06T22:41:53.418Z:
% 
% Does the WRBN count as sequence discriminative modeling? We still use a cross-entropy criterion for training the senone classifier...
%
% ^ <deblinbagchi@gmail.com> 2018-07-16T14:47:15.130Z:
% 
% How about sequence modeling instead of sequence-discriminative modeling ?
%
% ^ <deblinbagchi@gmail.com> 2018-07-16T14:47:44.943Z.

During evaluation, a forward-pass through the residual spectral mapper generates denoised features which are then fed to an off-the-shelf Kaldi recipe \cite{povey2011kaldi}. These features achieve a much lower WER on their own as compared to DNN spectral mappers trained without mimic loss~\cite{han2015deep,bagchi2015combining}. With the addition of the stronger feedback from a senone classifier, we achieve results beating the state-of-the-art system, which includes both additional noise-robust features, and joint training of the front-end denoiser with the acoustic model back-end.

\section{Prior Work}
\label{sec:prior}

For the task of robust ASR, there has been some attention paid to strategies such as adding noise-robust features to acoustic models \cite{wang2016joint}, using augmented training data \cite{du2016ustc}, and recurrent neural network language models \cite{du2016ustc,yoshioka2015ntt}. Another approach is to use a more sophisticated acoustic model, such as Convolutional Neural Networks (CNNs) \cite{qian2016very,zhang2017very}, Recurrent Neural Networks (RNNs) \cite{chen2015integration}, and Residual Memory Networks (RMNs) \cite{baskar2017residual} that use residual connections with DNNs.

In terms of front-end models, DNNs are the most common approach \cite{bagchi2015combining}, though RNNs have been used for speech enhancement as well, as in \cite{weninger2015speech}. There have also been a few studies that used CNNs for front-end speech denoising \cite{hui2015convolutional,fu2016snr,park2017fully}. In the last of these, the authors used a single "bypass" connection from the encoder to the decoder, but none of the models described here can be said to use residual connections. In addition, none of these authors evaluated the output of their model for the task of ASR.

Residual networks have seen success in computer vision~\cite{he2016deep,zagoruyko2016wide}, and speech recognition \cite{xiong2017microsoft,jahn2016wide}. These networks add shortcut connections to a neural network that pass the output of some layers to higher layers. The shortcut connections allow the network to compute a modification of the input, called the \textit{residual}, rather than having to re-compute the important parts of the input at every layer. This model seems a natural fit for the task of spectral mapping, which seeks to reproduce the input with the noise removed. We use an architecture similar to Wide ResNet with a small change: convolutional (channel-wise) dropout rather than conventional dropout. Architectural details are in Section~\ref{sec:architecture}.

Senone classification in speech recognition systems has improved due to recurrent neural networks. The horizontal connections in LSTMs work well in modeling the temporal nature of speech. On the other hand, convolutional neural networks are good for extracting useful patterns from spectral features. DNNs further complement the performance of these models by warping the speech manifold so that it resembles the senone feature space. The CNN-LSTM-DNN combination (CLDNN) along with HMMs have seen good results ~\cite{sainath2015convolutional,sainath2015learning}. Recently wide residual networks have been adapted for noise-robust speech recognition in the CHiME-4 setting and used with LSTMs and DNNs. This network, called WRBN is reported by~\cite{jahn2016wide} as a great acoustic model.

Mimic loss, proposed in~\cite{bagchi2018spectral}, is a kind of knowledge transfer that uses an acoustic model trained on clean speech to teach the speech enhancement model how to produce more realistic speech --- key to this idea is that the denoised speech should make a senone classifier {\em behave} like it is operating with clean speech. In contrast to joint training, the mimic loss does not tie the speech enhancement model to the particular acoustic model used; the enhancement module can be decoupled and used as a pure pre-processing unit with another recognizer. More details can be found in Section~\ref{sec:loss}.

\begin{figure}[t!]
\centering
\tikzstyle{block} = [rectangle, draw, minimum height=3.5em, minimum width=5em,align=center,fill=blue!10]
\tikzstyle{conv} = [block, fill=white, minimum width=7em, rounded corners]
\begin{tikzpicture}
	\node (input) {Noisy frames};
    \node[block, below=1em of input] (block1) {128 filter \\ block};
    \node[block, below=1em of block1] (block2) {128 filter \\ block};
    \node[block, below=1em of block2] (block3) {256 filter \\ block};
    \node[block, below=1em of block3] (block4) {256 filter \\ block};
    \node[block, below=1em of block4] (full) {Fully \\ connected};
    \node[below=1em of full] (output) {Denoised frame};
    
    \draw[-stealth,thick] (input) -- (block1);
    \draw[-stealth,thick] (block1) -- (block2);
    \draw[-stealth,thick] (block2) -- (block3);
    \draw[-stealth,thick] (block3) -- (block4);
    \draw[-stealth,thick] (block4) -- (full);
    \draw[-stealth,thick] (full) -- (output);
    
    \node[block, right=3em of block3, minimum height=18.2em, minimum width=10em] (outset) {};
    
    \draw[thick] (block3.north east) -- (outset.north west);
    \draw[thick] (block3.south east) -- (outset.south west);
    
    \node[conv, below=1em of outset.north] (conv1) {Size: $3 \times 3$ \\ Stride: $2 \times 2$};
    \node[conv, below=2em of conv1] (conv2) {Size: $3 \times 3$ \\ Stride: $1 \times 1$};
    \node[conv, below=1em of conv2] (conv3) {Size: $3 \times 3$ \\ Stride: $1 \times 1$};
    \node[circle, draw, below=1em of conv3, fill=white] (add) {+};
    
    \draw[-stealth,thick] (outset.north) -- (conv1);
    \draw[-stealth,thick] (conv1) -- (conv2);
    \draw[-stealth,thick] (conv2) -- (conv3);
    \draw[-stealth,thick] (conv3) -- (add);
    \draw[-stealth,thick] (add) -- (outset.south);
    
    \coordinate[below=0.9em of conv1] (a);
    \coordinate[right=4.3em of a] (b);
    \coordinate[right=4.3em of add.center] (c);
	\draw[-stealth,thick, rounded corners] (a) -- (b) -- (c) -- (add);
\end{tikzpicture}

\caption{Our residual network architecture consists of four blocks and two fully-connected layers. Each block starts with a convolutional layer for down-sampling and increasing the number of filters. The output of this block is used twice, once as input to the two convolutional layers that compute the residual, and again as the original signal that is modified by adding the computed residual.}
\label{fig:resnet}
\end{figure}
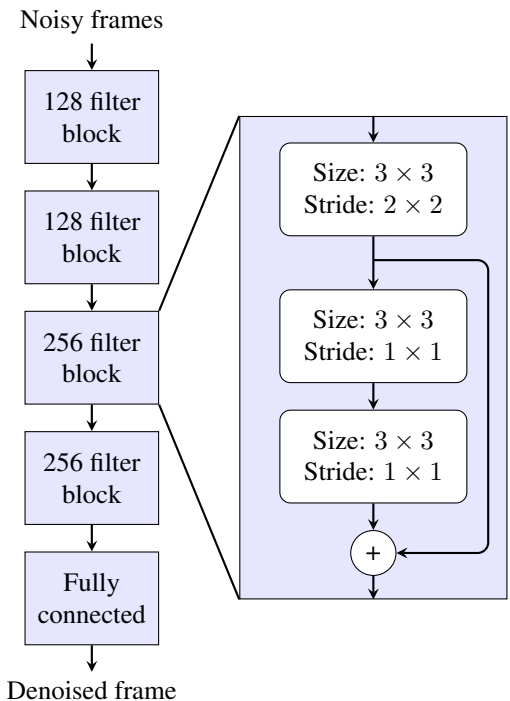

\section{Spectral Mapping}
\label{sec:mapping}

Spectral mapping improves performance of the speech recognizer by learning a mapping from noisy spectral patterns to clean ones. In our previous work \cite{han2015deep,bagchi2015combining}, we have shown that a DNN spectral mapper, which takes noisy spectrogram as input to predict clean filterbank features for ASR, yields good results on the CHiME-2 noisy and reverberant dataset. Specifically, we first divide the input time-domain signals into 25 ms windows with a 10 ms shift, and then apply short time Fourier transform (STFT) with a hamming window to compute log spectral magnitudes in each time frame. For a 16 kHz signal, each window contains 400 samples, and we use 512-point Fourier transform to compute the magnitudes, forming a 257-dimensional log magnitude vector. 

Many speech recognition systems extend the input features using delta and double-delta.  These features are a simple arithmetic function of the surrounding frames. CNNs naturally learn filters of a similar nature to the delta function, and can easily learn to approximate these features if necessary. We find that the model works better without these redundant features. We use 5 frames of stacked content (both past \& future) for both DNN and ResNets. Hence, the input feature dimension decreases to 2827 ($257 \times 11$) compared to 8481 when delta features are included ($257 \times 3 \times 11$).

%Specifically, we first divide the input time-domain signals into 25 ms frames with a 10 ms frame shift, and then apply short time Fourier transform (STFT) to compute log spectral magnitudes in each time frame. For a 16 kHz signal, each frame contains 400 samples, and we use 512-point Fourier transform to compute the magnitudes, forming a 257-dimensional log magnitude vector.

%Many speech recognition systems extend the input features using delta and double-delta.  These features are a simple arithmetic function of the surrounding frames. CNNs naturally learn filters of a similar nature to the delta function, and can easily learn to approximate these features if necessary. We find that the model works better without these redundant features.

\subsection{Baseline Model}
\label{sec:baseline}

We use a baseline model for comparison, a DNN that is also a front-end-only system. Though this architecture is quite a bit simpler than the residual network architecture, similar architectures are commonly used in speech enhancement research~\cite{han2015deep,wang2016joint}. 

Unlike the proposed model, we add delta and double-delta features to the input for the baseline model, since a DNN cannot learn the delta function as easily. These features have been shown to dramatically improve ASR performance, and they improve spectral mapping performance as well.

Our baseline model is a 2-layer DNN with 2048 ReLU neurons in each layer, with an output layer of 257 neurons. We use batch norm and dropout to regularize the network. The batch norm uses the moving mean and variance at training time as well as test time. This is the same architecture that is used in~\cite{bagchi2018spectral}.

\subsection{ResNet Architecture}
\label{sec:architecture}

A residual network adds shortcut connections to neural network architectures, typically CNNs, in a way that causes the network to learn a modification of the original input, rather than being forced to reconstruct the important information at each layer. This usually takes the form of blocks of several neural network layers with the output of the first layer added to the output of the last layer, so that the interior layers can compute the residual.

Adding these connections has several advantages: the training time is decreased, the networks can grow deeper, and the model tends to behave more like an ensemble of smaller models~\cite{veit2016residual}. In addition to all of those, however, we expect this model to be particularly good for the task of speech denoising, since the architecture matches the task at hand: reconstructing the input signal with the residual noisy signal removed.

\begin{figure*}[t]
\tikzstyle{model} = [rectangle, draw, minimum height=4em, minimum width=5em,align=center]
\centering
\begin{tikzpicture}
	\node (a) at (0,0) {\large \textbf{Pretraining:}};
	\node[below=3em of a.south west,anchor=west,align=center] (crit_in) {Clean \\ speech};
    \node[model, right=1em of crit_in] (crit) {Senone \\ classifier};
    \node[right=1em of crit,align=center,minimum height=2em] (crit_out) {Predicted \\ senones};
    \node[above right=0em and 4em of crit_out, align=center, minimum height=2em] (senone) {Senone \\ label};
    
    \draw[-stealth,thick] (crit_in) -- (crit);
    \draw[-stealth,thick] (crit) -- (crit_out);
    \draw[stealth-stealth,thick] (crit_out) -- node[below right,pos=0.3] {\large $\mathcal{L}_{\rm Cross-entropy}$} (senone);
    
    \node[below=3em of crit_in, align=center] (act_in) {Noisy \\ speech};
    \node[model, right=1em of act_in] (act) {Spectral \\ mapper};
    \node[right=1em of act,align=center,minimum height=2em] (act_out) {Denoised \\ speech};
    \node[above right=0em and 4em of act_out,align=center,minimum height=2em] (clean) {Clean \\ speech};

	\draw[-stealth,thick] (act_in) -- (act);
    \draw[-stealth,thick] (act) -- (act_out);
    \draw[stealth-stealth,thick] (act_out) -- node[below right,pos=0.3] {\large $\mathcal{L}_{\rm Fidelity}$} (clean);
    
    \node[right=17em of a] (c) {\large \textbf{Mimic loss training:}};
    
    \node[below=9em of c.south west,anchor=west,align=center] (joint_noisy) {Noisy \\ speech};
    \node[model, right=1em of joint_noisy] (joint_actor) {Spectral \\ mapper};
    \node[right=1em of joint_actor,align=center,minimum height=3em,minimum width=5em] (joint_act_out) {Denoised \\ speech};
    \node[above=4em of joint_act_out,align=center, minimum height=3em, minimum width=5em] (joint_clean) {Clean \\ speech};
    
    \node[model, right=1em of joint_clean, fill=gray!30] (output_critic) {Senone \\ classifier};
    \node[right=1em of output_critic,align=center,minimum height=3em, minimum width=5em] (soft_labels) {Soft \\ labels};
    
    \node[model, right=1em of joint_act_out, fill=gray!30] (joint_critic) {Senone \\ classifier};
    \node[right=1em of joint_critic,align=center,minimum height=3em,minimum width=5em] (joint_crit_out) {Classifier \\ output};
    \draw[-stealth,thick] (joint_noisy) -- (joint_actor);
    \draw[stealth-stealth,thick] (joint_act_out) -- node[left,pos=0.5] {\large $\mathcal{L}_{\rm Fidelity}$} (joint_clean);
    \draw[-stealth,thick] (joint_clean) -- (output_critic);
    \draw[-stealth,thick] (output_critic) -- (soft_labels);
    \draw[-stealth,thick] (joint_actor) -- (joint_act_out);
    \draw[-stealth,thick] (joint_act_out) -- (joint_critic);
    \draw[-stealth,thick] (joint_critic) -- (joint_crit_out);
    \draw[stealth-stealth,thick] (joint_crit_out) -- node[left] {\large $\mathcal{L}_{\rm Mimic}$} (soft_labels);

\end{tikzpicture}
\caption{When using mimic loss, the enhancement system is trained in two stages. In the pretraining stage, the senone classifier is trained on clean speech to predict senone labels with cross-entropy criterion ($\mathcal{L}_{\rm Cross-entropy}$) and the spectral mapper is pretrained to map from noisy speech to clean speech using MSE criterion (fidelity loss, $\mathcal{L}_{\rm Fidelity}$). In the mimic loss training stage, the pretrained spectral mapper is trained further using both fidelity loss and mimic loss ($\mathcal{L}_{\rm Mimic}$), the loss between the two sets of outputs from the classifier when fed parallel clean and denoised utterances. The gray models have frozen weights.}
\label{fig:systemdesc}
\end{figure*}
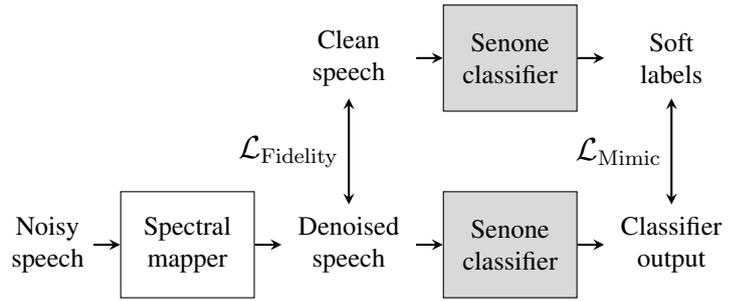

In previous work using CNNs for speech enhancement, it has been noted that performance sometimes degrades with the addition of max pooling between convolution layers \cite{fu2016snr}. We also observe this phenomenon, and instead of doing max pooling, we use an additional CNN layer with stride $2 \times 2$ to learn a down-sampling function. This layer has the additional effect of increasing the number of filters, so the output can be directly added to the output of the last layer of the block as a residual connection.

Inspired by Wide ResNet~\cite{zagoruyko2016wide}, we use dropout instead of batch normalization, though we use convolutional (channel-wise) dropout rather than conventional dropout in order to better preserve the local structure within each filter. This results in a small gain in WER of around 0.2 percent. The authors also suggested that a shallower, but wider network may work better than a very deep network; we use a network that is only 14 layers deep, and the layers are a comparable width to Wide ResNet. Neither adding filters nor layers improved performance.

The full architecture of this model can be seen in Figure~\ref{fig:resnet}. The first part of the model uses four convolutional blocks: two blocks of 128 filters, and two of 256 filters. After the convolutional blocks, we append two fully-connected layers of 2048 neurons and an output layer. The whole network uses ReLU neurons.

\subsection{Training the Residual Network}
\label{sec:training}

We found that there was some sensitivity to the training procedure for the residual network, so we report our procedure here. We use the Adam optimizer~\cite{kingma2014adam} with an initial learning rate of $10^{-4}$ and an exponential decay rate of 0.95 every $10^4$ steps. Following the training procedure for ResNets in the field of computer vision, we experimented with learning rate drops, going to one-tenth of the initial learning rate after convergence. We found that this resulted in sizable improvements in the fidelity loss (see Table~\ref{tab:mse} in Section~\ref{sec:results}), but no improvements in the final WER.

Training a model to faithfully reproduce the input via fidelity loss does not teach a model exactly what parts of the signal are important to focus on reproducing correctly. A lower learning rate allows the model to make more precise adjustments to its parameters, reproducing small details in the spectrogram more faithfully. However, the fact that these details don't help for the task of speech recognition indicates that they are mostly irrelevant for speech comprehension.

\section{Mimic Loss Training}
\label{sec:loss}

In order to train with mimic loss, first the two component models must be pre-trained. We pre-train the spectral mapper to compute the function $f(\cdot)$ from a noisy spectral component $x^k_m$ for frequency $k$ at time slice $m$, augmented with a five-frame window (designated $\tilde{x}_m^k=[x_{m\pm 5}^k]$), to clean spectral slice $y_m$. This is called the \textit{fidelity loss}, written as follows:

\begin{equation}\label{eqn1}
\begin{split}
 \mathcal{L}_{\rm Fidelity}(\tilde{x}_m,y_m) & = \frac{1}{K}\sum_{k=1}^{K} (y_m^k - f(\tilde{x}_m^k))^{2}  \\
\end{split}
\end{equation}

\bigskip

\noindent While the residual network spectral mapper trained with only fidelity loss results in performance better than previous front-end-only systems, we add \textit{mimic loss} for an additional gain in performance. This is done by training a senone classifier to learn a function $g(\cdot)$ from clean speech input $\tilde{y}_m$ to a set of $D$ senones, and freezing the weights of the model. The spectral mapper is then trained to mimic the behavior of clean speech by backpropagating the L2-loss between clean and denoised input after being run through the acoustic model. The loss is computed at the output layer, before softmax is applied.

\begin{equation}\label{eqn2}
\begin{split}
 \mathcal{L}_{\rm Mimic}(\tilde{\tilde{x}}_m, \tilde{y}_m) = \frac{1}{D}\sum_{d=1}^{D} (g(\tilde{y}_m)^d - g(\tilde{f}(\tilde{x}_m))^d)^2
\end{split}
\end{equation}

\bigskip

\noindent In early experiments, we found that using only mimic loss did not allow the model to converge, since it cares only about behavior and not the actual shape of the features. So we use a linear combination of fidelity loss and mimic loss:

\begin{equation}\label{joint_loss}
\begin{split}
 \mathcal{L}_{\rm Joint} = \mathcal{L}_{\rm Fidelity} + \alpha \mathcal{L}_{\rm Mimic}
\end{split}
\end{equation}

\bigskip

\noindent where $\alpha$ is a hyper-parameter controlling the ratio of fidelity and mimic losses. For our experiments, we use $\alpha=0.1$ when the mimic model is a DNN, and $\alpha=0.05$ when the mimic model is a WRBN. These values were chosen to ensure that the magnitude of the fidelity loss and mimic loss were roughly equal. Higher or lower values of $\alpha$ do not usually produce better results. The entire process for training with mimic loss can be seen in Figure~\ref{fig:systemdesc}.

\subsection{Senone Classification}

In order to provide additional feedback to the spectral mapper, we train another model as a teaching model. This second model is trained for the task of senone classification, with clean speech as input. This model will ideally learn what parts of a speech signal are important for recognition and be able to help a spectral mapper model to learn to reproduce these important speech structures faithfully.

The loss used to train the senone classifier is typical acoustic model criterion: the cross-entropy loss between the outputs of the classifier, $g(\tilde{y}_m)$, and the senone label, $z_m$, where $g(\cdot)$ is the function computed by the classifier.

\subsection{Senone Classifier Models}

%\begin{table}[t!]
 % \caption{Development set cross-entropy loss for the senone classifier models trained on clean speech.}
 % \label{tab:classification}
 % \centering
%  \vspace{0.5cm}
 % \begin{tabular}{ l c c }
 %   \toprule
%    Senone Classification Model          & CE Loss \\
%    \midrule
%    Fully-connected network       & 2.1     \\
%    Wide Residual BiLSTM           & 1.1     \\
%    \bottomrule
%  \end{tabular}
  
%\end{table}

We experiment with two different senone classifier models that are separate from the one used in the off-the-shelf Kaldi recipe used for recognition. This separation exists both in terms of the architecture and particular parameter values that are used,  which gives some evidence to our claim that our front-end model is not tied to any particular acoustic model. For both models, we target 1999 senone classes.

For our first model, we use a 6-layer 1024-node DNN with batch norm and leaky ReLU neurons (with a leak factor of 0.3), the same model used in \cite{bagchi2018spectral}. Our second model is a WRBN model that has recently been shown to perform well on the CHiME-4 challenge~\cite{jahn2016wide}. This allows us to add a sequential component to the training of the residual network via the senone classifier.

The WRBN model combines a wide residual network to a bi-directional LSTM model. The wide residual network consists of 3 residual blocks of 6 convolutional layers each, with 80, 160, and 320 channels. The first layer in the second and third blocks use a stride of 2 $\times$ 2 to downsample, with a 1 $\times$ 1 convolutional layer bypass connection. Following these blocks is a linear layer.

The LSTM part of the model is a 2-layer network with 512 nodes per layer in each direction. After the first layer, the two directions are added together before being passed to the second layer, after which the two directions are concatenated. The last two layers in the network are linear. The entire network uses ELU activations~\cite{clevert2015fast}, batch norm, and dropout.

Both the classifier networks are trained using the Adam optimizer~\cite{kingma2014adam} with learning rate $\eta = 10^{-4}$ for the WRBN and $\eta=10^-5$ for the DNN. We use 257 dimensional mean-normalized spectrogram features as input to the networks. Delta and delta-delta coefficients are not used. The DNN senone classifier uses a window of 5 context frames in the past and future while the WRBN is trained on a per utterance basis with full backpropagation through time. 

The WRBN model achieved a cross-entropy loss of 1.1 on the clean speech development set, which is almost half the cross-entropy loss of the DNN model, which was 2.1, so we expect it to be able to provide much more helpful feedback to the spectral mapper model.

\section{Experiments}
\label{sec:experiments}

\begin{table}[t!]
  \caption{Fidelity loss on the development set for our baseline model and the residual network, both with and without mimic loss training.}
  \label{tab:mse}
  \centering
  \vspace{0.5cm}
  \begin{tabular}{ l c c }
    \toprule
	%Enhancement Model         & Parameters & Fidelity loss \\
	Enhancement Model          & Fidelity loss \\
    \midrule
    %Fully-connected network      & 22.1 M  & 0.52     \\
    %\qquad with DNN mimic       &         & 0.51     \\
    %\qquad with WRBN mimic       &         & 0.51     \\
    DNN spectral mapper      & 0.52     \\
    \qquad with DNN mimic        & 0.51     \\
    \qquad with WRBN mimic        & 0.51     \\
    %ResNet-style network         & 17.6 M  & 0.47     \\
    %\qquad with learn rate drop  &         & 0.44     \\
    %\qquad with DNN mimic       &         & 0.48     \\
    %\qquad with WRBN mimic       &         & 0.49     \\
    Residual network mapper           & 0.47     \\
    \qquad with learn rate drop   & 0.44     \\
    \qquad with DNN mimic        & 0.48     \\
    \qquad with WRBN mimic       & 0.49     \\
    \bottomrule
  \end{tabular}
  
\end{table}

We evaluate the quality of the denoised features produced with our residual network spectral mapper by training an off-the-shelf Kaldi recipe for Track 2 of the CHiME-2 challenge~\cite{vincent2013second}.

\subsection{Task and data description}
\label{sec:task}

CHiME-2 is a medium-vocabulary task for word recognition under reverberant and noisy environments without speaker movements. In this task, three types of data are provided based on the Wall Street Journal (WSJ0) 5K vocabulary read speech corpus: clean, reverberant and reverberant+noisy. The clean utterances are extracted from the WSJ0 database. The reverberant utterances are created by convolving the clean speech with binaural room impulse responses (BRIR) corresponding to a frontal position in a family living room. Real-world non-stationary noise background recorded in the same room is mixed with the reverberant utterances to form the reverberant+noisy set. The noise excerpts are selected such that the signal-to-noise ratio (SNR) ranges among -6, -3, 0, 3, 6 and 9 dB without scaling. The multi-condition training, development and test sets of the reverberant+noisy set contain 7138, 2454 and 1980 utterances respectively, which are the same utterances in the clean set but with reverberation and noise at 6 different SNR conditions.

\subsection{Description of the Kaldi recipe}
\label{sec:acoustic}

In order to determine the effectiveness of our front-end system, we train the denoised features with an off-the-shelf Kaldi recipe for CHiME-2. The DNN-HMM hybrid system is trained using the clean WSJ0-5k alignments generated using the method stated above. The DNN acoustic model has 7 hidden layers, with 2048 sigmoid neurons in each layer and a softmax output layer. Splicing context size for the filterbank features was fixed at 11 frames (5 frames of past and 5 frames of future context), with the minibatch-size being 1024. After that, we train the DNN with state-level minimum Bayes risk (sMBR) sequence training. We regenerate the lattices after the first iteration and train for 4 more iterations. We use the CMU pronunciation dictionary and the official 5k closed-vocabulary trigram language model in our experiments.

\section{Results}
\label{sec:results}

\begin{table}[t!]
  \caption{Word error rates after generating denoised features and feeding them to off-the-shelf Kaldi recipe for training.  The first line for each model indicates WER for models trained with fidelity loss only; the second includes the joint fidelity-mimic loss.}
  \label{tab:wer}
  \centering
  \vspace{0.5cm}
  \begin{tabular}{l c c}
    \toprule
    %Enhancement Model        & Parameters & WER \\
    Enhancement Model         & WER \\
    \midrule
    %No enhancement           &            & 17.3 \\
    %DNN spectral mapper      & 22.1 M     & 16.5\\
    %\qquad with DNN mimic     &            & 14.4 \\
    %\qquad with WRBN mimic    &            & 14.3 \\
    %Residual network mapper   & 17.6 M     & 11.1 \\
    %\qquad with DNN mimic     &            & 10.7 \\
    %\qquad with WRBN mimic    &            & 10.1 \\
    No enhancement           & 17.3 \\
    DNN spectral mapper      & 16.0\\
    \qquad with DNN mimic    & 14.4 \\
    \qquad with WRBN mimic   & 14.0 \\
    Residual network mapper  & 10.8 \\
    \qquad with DNN mimic    & 10.5 \\
    \qquad with WRBN mimic   & 9.3 \\
    \bottomrule
  \end{tabular}
\end{table}

We report the best fidelity loss of all models on the development set in  Table~\ref{tab:mse}. Fidelity loss is a record of how well a model can exactly reproduce the clean speech signal, not taking into account whether the denoised signal is speech-like or not. In terms of fidelity loss, our residual networks gain about 10\% over the baseline models. With the learning rate drop that is common in vision tasks, residual networks gain an additional 5\%. However, this improvement in fidelity loss did not translate to any gain in WER. The last entries in the table show that the residual network performs slightly worse in terms of fidelity loss when mimic is added, which is to be expected given that the objective is split between fidelity loss and mimic loss.

In addition to our fidelity loss results, we present robust speech recognition results, generated by presenting our denoised spectral features to an off-the-shelf Kaldi recipe. The results are shown in Table~\ref{tab:wer}. One point of note is that the features generated by the DNN spectral mapper without mimic loss only perform a little better than the original noisy features, likely due to introduced distortions~\cite{narayanan2014investigation}.

It is also interesting to note that the WER gain for the residual network is much more significant than the fidelity loss alone would suggest, reaching around 30\% relative improvement. This improvement holds whether the model is trained with or without mimic loss. Finally, we note that using a more sophisticated WRBN mimic leads to a large improvement in the performance of the residual network spectral mapper, but only a small gain for the DNN mapper. We speculate that the modeling power of the DNN may be limited, since it has only two layers. 
% * <plantinga.peter@protonmail.com> 2018-07-16T16:49:28.201Z:
% 
% > due to introduced distortions
% Citation needed?
% 
% ^ <plantinga.peter@protonmail.com> 2018-07-16T19:19:20.701Z.

Finally, we compare our best-performing model with other studies on the CHiME-2 test set that use only feature engineering and generation (e.g. more sophisticated language models not included). Even without mimic loss, our model performs much better than all other systems that use no additional noise-robust features or joint training of front-end speech enhancer and acoustic model. With the addition of mimic loss, our model also performs 10\% better than the state-of-the-art, which uses both of these.

\begin{table}[t!]
  \caption{Performance comparison with other studies on the CHiME2 test set.  ``Additional NR features'' indicates that noise-robust features are added.  ``Joint ASR training'' indicates that the final ASR system and enhancement model are jointly tuned.  Our previous system \cite{bagchi2018spectral} was a DNN trained with joint fidelity-mimic loss. }
  \label{tab:comparisons}
  
  \addtolength{\tabcolsep}{-2px}
  \centering
  \vspace{0.5cm}
  \begin{tabular}{ l c c c c }
    \toprule
         & Additional & Joint ASR & \\
    Study & NR features & training & WER \\
    \midrule
    Chen et. al \cite{chen2015integration}       & - & \checkmark & 16.0 \\
    Narayanan-Wang \cite{narayanan2015improving} & \checkmark & \checkmark & 15.4 \\
    Bagchi et. al \cite{bagchi2018spectral}      & - & - & 14.7 \\
    Weninger et.al \cite{weninger2015speech}     & \checkmark & - & 13.8 \\
    Wang et.al \cite{wang2016joint}              & \checkmark & \checkmark & 10.6 \\
    \midrule
    Residual network                         & -  & -  & 10.8 \\
    \textbf{ResNet + mimic loss}                     & - & - & \textbf{9.3} \\
    \bottomrule
  \end{tabular}
\end{table}

\section{Conclusions}

We have enhanced the performance of the mimic loss framework with the help of a ResNet-style architecture for spectral mapping and a more sophisticated senone classifier, with an almost 30\% improvement over the DNN baseline and achieve the best acoustic-only adaptation result without using additional noise-robust features or joint training of a speech enhancement module and ASR system.

%In future, we hope to use mimic loss where the outputs of multiple layers of the acoustic model are compared, instead of just the output layer. Preliminary results suggest that the higher layers are more informative than the lower layers, and including one or two additional layers to the loss can help performance.

%We also hope to investigate the limited improvement in the WER of the DNN mapper when using a sophisticated WRBN mimic model. One route to address this is to try using many mimic models to see if any of them can improve on the DNN mimic model.

One route to achieving improved WER may be to do mimic loss at a higher level, such as the word level rather than the senone level. Since other work has found that joint training all the way up to the word level has helped performance, we expect that this would help our denoiser.

For some tasks, targeting an ideal ratio mask which is then multiplied with the original signal has achieved higher performance than spectral mapping. We plan to apply mimic loss to the technique of spectral masking; if successful, we could extend our work to the CHiME-3 and CHiME-4 challenges where mask generation during the beamforming stage has achieved the state-of-the-art.

%Finally, we hope to remove the dependence in our model on artificially generated parallel clean-noisy data by using beamformed data instead of clean data. Perhaps such a system using mimic loss could learn to do a beamforming-like mapping on a single channel. 

Our code is publicly available at \url{https://github.com/OSU-slatelab/residual_mimic_net}.

\section{Acknowledgements}

This work was supported by the National Science Foundation under Grant IIS-1409431. We also thank the Ohio Supercomputer Center (OSC) \cite{OhioSupercomputerCenter1987} for providing us with computational resources.  We gratefully acknowledge the support of NVIDIA Corporation with the donation of the Quadro P6000 GPU used for this research.

\bibliographystyle{IEEEbib}

\bibliography{mybib}

\begin{thebibliography}{10}

\bibitem{bagchi2018spectral}
Deblin Bagchi, Peter Plantinga, Adam Stiff, and Eric Fosler-Lussier,
\newblock ``Spectral feature mapping with mimic loss for robust speech
  recognition,''
\newblock in {\em Audio, Speech, and Signal Processing (ICASSP), International
  Conference on}, 2018.

\bibitem{narayanan2015improving}
Arun Narayanan and DeLiang Wang,
\newblock ``Improving robustness of deep neural network acoustic models via
  speech separation and joint adaptive training,''
\newblock {\em IEEE/ACM Transactions on Audio, Speech, and Language
  Processing}, vol. 23, no. 1, pp. 92--101, 2015.

\bibitem{han2014learning}
Kun Han, Yuxuan Wang, and DeLiang Wang,
\newblock ``Learning spectral mapping for speech dereverberation,''
\newblock in {\em Acoustics, Speech and Signal Processing (ICASSP), 2014 IEEE
  International Conference on}. IEEE, 2014, pp. 4628--4632.

\bibitem{han2015learning}
Kun Han, Yuxuan Wang, DeLiang Wang, William~S Woods, Ivo Merks, and Tao Zhang,
\newblock ``Learning spectral mapping for speech dereverberation and
  denoising,''
\newblock {\em IEEE Transactions on Audio, Speech, and Language Processing},
  vol. 23, no. 6, pp. 982--992, 2015.

\bibitem{han2015deep}
Kun Han, Yanzhang He, Deblin Bagchi, Eric Fosler-Lussier, and DeLiang Wang,
\newblock ``Deep neural network based spectral feature mapping for robust
  speech recognition,''
\newblock in {\em Sixteenth Annual Conference of the International Speech
  Communication Association}, 2015.

\bibitem{bagchi2015combining}
Deblin Bagchi, Michael~I Mandel, Zhongqiu Wang, Yanzhang He, Andrew Plummer,
  and Eric Fosler-Lussier,
\newblock ``Combining spectral feature mapping and multi-channel model-based
  source separation for noise-robust automatic speech recognition,''
\newblock in {\em Automatic Speech Recognition and Understanding (ASRU), 2015
  IEEE Workshop on}. IEEE, 2015, pp. 496--503.

\bibitem{wang2016joint}
Zhong-Qiu Wang and DeLiang Wang,
\newblock ``A joint training framework for robust automatic speech
  recognition,''
\newblock {\em IEEE/ACM Transactions on Audio, Speech, and Language
  Processing}, vol. 24, no. 4, pp. 796--806, 2016.

\bibitem{ba2014deep}
Jimmy Ba and Rich Caruana,
\newblock ``Do deep nets really need to be deep?,''
\newblock in {\em Advances in neural information processing systems}, 2014, pp.
  2654--2662.

\bibitem{hinton2015distilling}
Geoffrey Hinton, Oriol Vinyals, and Jeff Dean,
\newblock ``Distilling the knowledge in a neural network,''
\newblock {\em arXiv preprint arXiv:1503.02531}, 2015.

\bibitem{lopez2015unifying}
David Lopez-Paz, L{\'e}on Bottou, Bernhard Sch{\"o}lkopf, and Vladimir Vapnik,
\newblock ``Unifying distillation and privileged information,''
\newblock {\em arXiv preprint arXiv:1511.03643}, 2015.

\bibitem{li2014learning}
Jinyu Li, Rui Zhao, Jui-Ting Huang, and Yifan Gong,
\newblock ``Learning small-size {DNN} with output-distribution-based
  criteria,''
\newblock in {\em Fifteenth annual conference of the international speech
  communication association}, 2014.

\bibitem{he2016deep}
Kaiming He, Xiangyu Zhang, Shaoqing Ren, and Jian Sun,
\newblock ``Deep residual learning for image recognition,''
\newblock in {\em Proceedings of the IEEE conference on computer vision and
  pattern recognition}, 2016, pp. 770--778.

\bibitem{jahn2016wide}
Lukas~Drude Jahn~Heymann and Reinhold Haeb-Umbach,
\newblock ``Wide residual {BLSTM} network with discriminative speaker
  adaptation for robust speech recognition,''
\newblock in {\em Proceedings of the 4th International Workshop on Speech
  Processing in Everyday Environments (CHiME’16)}, 2016, pp. 12--17.

\bibitem{graves2013speech}
Alex Graves, Abdel-rahman Mohamed, and Geoffrey Hinton,
\newblock ``Speech recognition with deep recurrent neural networks,''
\newblock in {\em Acoustics, speech and signal processing (icassp), 2013 ieee
  international conference on}. IEEE, 2013, pp. 6645--6649.

\bibitem{graves2013hybrid}
Alex Graves, Navdeep Jaitly, and Abdel-rahman Mohamed,
\newblock ``Hybrid speech recognition with deep bidirectional {LSTM},''
\newblock in {\em Automatic Speech Recognition and Understanding (ASRU), 2013
  IEEE Workshop on}. IEEE, 2013, pp. 273--278.

\bibitem{povey2011kaldi}
Daniel Povey, Arnab Ghoshal, Gilles Boulianne, Lukas Burget, Ondrej Glembek,
  Nagendra Goel, Mirko Hannemann, Petr Motlicek, Yanmin Qian, Petr Schwarz, Jan
  Silovsky, Georg Stemmer, and Karel Vesely,
\newblock ``The kaldi speech recognition toolkit,''
\newblock in {\em IEEE 2011 Workshop on Automatic Speech Recognition and
  Understanding}. Dec. 2011, IEEE Signal Processing Society,
\newblock IEEE Catalog No.: CFP11SRW-USB.

\bibitem{du2016ustc}
Jun Du, Yan-Hui Tu, Lei Sun, Feng Ma, Hai-Kun Wang, Jia Pan, Cong Liu,
  Jing-Dong Chen, and Chin-Hui Lee,
\newblock ``The {USTC-iFlytek} system for {CHiME-4} challenge,''
\newblock {\em Proc. CHiME}, pp. 36--38, 2016.

\bibitem{yoshioka2015ntt}
Takuya Yoshioka, Nobutaka Ito, Marc Delcroix, Atsunori Ogawa, Keisuke
  Kinoshita, Masakiyo Fujimoto, Chengzhu Yu, Wojciech~J Fabian, Miquel Espi,
  Takuya Higuchi, et~al.,
\newblock ``The {NTT} {CHiME-3} system: Advances in speech enhancement and
  recognition for mobile multi-microphone devices,''
\newblock in {\em Automatic Speech Recognition and Understanding (ASRU), 2015
  IEEE Workshop on}. IEEE, 2015, pp. 436--443.

\bibitem{qian2016very}
Yanmin Qian and Philip~C Woodland,
\newblock ``Very deep convolutional neural networks for robust speech
  recognition,''
\newblock in {\em Spoken Language Technology Workshop (SLT), 2016 IEEE}. IEEE,
  2016, pp. 481--488.

\bibitem{zhang2017very}
Yu~Zhang, William Chan, and Navdeep Jaitly,
\newblock ``Very deep convolutional networks for end-to-end speech
  recognition,''
\newblock in {\em Acoustics, Speech and Signal Processing (ICASSP), 2017 IEEE
  International Conference on}. IEEE, 2017, pp. 4845--4849.

\bibitem{chen2015integration}
Z~Chen, S~Watanabe, H~Erdogan, and JR~Hershey,
\newblock ``Integration of speech enhancement and recognition using long-short
  term memory recurrent neural network,''
\newblock in {\em Proc. Interspeech}, 2015.

\bibitem{baskar2017residual}
Murali~Karthick Baskar, Martin Karafi{\'a}t, Luk{\'a}{\v{s}} Burget, Karel
  Vesel{\`y}, Franti{\v{s}}ek Gr{\'e}zl, and Jan {\v{C}}ernock{\`y},
\newblock ``Residual memory networks: Feed-forward approach to learn long-term
  temporal dependencies,''
\newblock in {\em Acoustics, Speech and Signal Processing (ICASSP), 2017 IEEE
  International Conference on}. IEEE, 2017, pp. 4810--4814.

\bibitem{weninger2015speech}
Felix Weninger, Hakan Erdogan, Shinji Watanabe, Emmanuel Vincent, Jonathan
  Le~Roux, John~R Hershey, and Bj{\"o}rn Schuller,
\newblock ``Speech enhancement with {LSTM} recurrent neural networks and its
  application to noise-robust {ASR},''
\newblock in {\em International Conference on Latent Variable Analysis and
  Signal Separation}. Springer, 2015, pp. 91--99.

\bibitem{hui2015convolutional}
Like Hui, Meng Cai, Cong Guo, Liang He, Wei-Qiang Zhang, and Jia Liu,
\newblock ``Convolutional maxout neural networks for speech separation,''
\newblock in {\em Signal Processing and Information Technology (ISSPIT), 2015
  IEEE International Symposium on}. IEEE, 2015, pp. 24--27.

\bibitem{fu2016snr}
Szu-Wei Fu, Yu~Tsao, and Xugang Lu,
\newblock ``{SNR}-aware convolutional neural network modeling for speech
  enhancement.,''
\newblock in {\em Proc. Interspeech}, 2016, pp. 3768--3772.

\bibitem{park2017fully}
Se~Rim Park and Jin~Won Lee,
\newblock ``A fully convolutional neural network for speech enhancement,''
\newblock {\em Proc. Interspeech 2017}, pp. 1993--1997, 2017.

\bibitem{zagoruyko2016wide}
Sergey Zagoruyko and Nikos Komodakis,
\newblock ``Wide residual networks,''
\newblock in {\em Proceedings of the British Machine Vision Conference (BMVC)},
  Edwin R.~Hancock Richard C.~Wilson and William A.~P. Smith, Eds. September
  2016, pp. 87.1--87.12, BMVA Press.

\bibitem{xiong2017microsoft}
Wayne Xiong, Jasha Droppo, Xuedong Huang, Frank Seide, Mike Seltzer, Andreas
  Stolcke, Dong Yu, and Geoffrey Zweig,
\newblock ``The {Microsoft} 2016 conversational speech recognition system,''
\newblock in {\em Acoustics, Speech and Signal Processing (ICASSP), 2017 IEEE
  International Conference on}. IEEE, 2017, pp. 5255--5259.

\bibitem{sainath2015convolutional}
Tara~N Sainath, Oriol Vinyals, Andrew Senior, and Ha{\c{s}}im Sak,
\newblock ``Convolutional, long short-term memory, fully connected deep neural
  networks,''
\newblock in {\em Acoustics, Speech and Signal Processing (ICASSP), 2015 IEEE
  International Conference on}. IEEE, 2015, pp. 4580--4584.

\bibitem{sainath2015learning}
Tara~N Sainath, Ron~J Weiss, Andrew Senior, Kevin~W Wilson, and Oriol Vinyals,
\newblock ``Learning the speech front-end with raw waveform {CLDNN}s,''
\newblock in {\em Sixteenth Annual Conference of the International Speech
  Communication Association}, 2015.

\bibitem{veit2016residual}
Andreas Veit, Michael~J Wilber, and Serge Belongie,
\newblock ``Residual networks behave like ensembles of relatively shallow
  networks,''
\newblock in {\em Advances in Neural Information Processing Systems}, 2016, pp.
  550--558.

\bibitem{kingma2014adam}
Diederik~P Kingma and Jimmy Ba,
\newblock ``Adam: A method for stochastic optimization,''
\newblock {\em arXiv preprint arXiv:1412.6980}, 2014.

\bibitem{clevert2015fast}
Djork-Arn{\'e} Clevert, Thomas Unterthiner, and Sepp Hochreiter,
\newblock ``Fast and accurate deep network learning by exponential linear units
  {(ELUs)},''
\newblock {\em arXiv preprint arXiv:1511.07289}, 2015.

\bibitem{vincent2013second}
Emmanuel Vincent, Jon Barker, Shinji Watanabe, Jonathan Le~Roux, Francesco
  Nesta, and Marco Matassoni,
\newblock ``The second {‘CHiME’} speech separation and recognition
  challenge: Datasets, tasks and baselines,''
\newblock in {\em Acoustics, Speech and Signal Processing (ICASSP), 2013 IEEE
  International Conference on}. IEEE, 2013, pp. 126--130.

\bibitem{narayanan2014investigation}
Arun Narayanan and DeLiang Wang,
\newblock ``Investigation of speech separation as a front-end for noise robust
  speech recognition,''
\newblock {\em IEEE/ACM Transactions on Audio, Speech, and Language
  Processing}, vol. 22, no. 4, pp. 826--835, 2014.

\bibitem{OhioSupercomputerCenter1987}
Ohio~Supercomputer Center,
\newblock ``Ohio supercomputer center,''
  \url{http://osc.edu/ark:/19495/f5s1ph73}, 1987.

\end{thebibliography}

% \begin{thebibliography}{9}
% \bibitem[1]{Davis80-COP}
%   S.\ B.\ Davis and P.\ Mermelstein,
%   ``Comparison of parametric representation for monosyllabic word recognition in continuously spoken sentences,''
%   \textit{IEEE Transactions on Acoustics, Speech and Signal Processing}, vol.~28, no.~4, pp.~357--366, 1980.
% \bibitem[2]{Rabiner89-ATO}
%   L.\ R.\ Rabiner,
%   ``A tutorial on hidden Markov models and selected applications in speech recognition,''
%   \textit{Proceedings of the IEEE}, vol.~77, no.~2, pp.~257-286, 1989.
% \bibitem[3]{Hastie09-TEO}
%   T.\ Hastie, R.\ Tibshirani, and J.\ Friedman,
%   \textit{The Elements of Statistical Learning -- Data Mining, Inference, and Prediction}.
%   New York: Springer, 2009.
% \bibitem[4]{YourName17-XXX}
%   F.\ Lastname1, F.\ Lastname2, and F.\ Lastname3,
%   ``Title of your INTERSPEECH 2017 publication,''
%   in \textit{Interspeech 2017 -- 18\textsuperscript{th} Annual Conference of the International Speech Communication Association, August 20?24, Stockholm, Sweden, Proceedings, Proceedings}, 2017, pp.~100--104.
% \bibitem[5]{BiRNN} Schuster, Mike, and Kuldip K. Paliwal. "Bidirectional recurrent neural networks." Signal Processing, IEEE Transactions on 45.11 (1997): 2673-2681.2.
% \end{thebibliography}

\end{document}